\documentclass[iop,apj]{emulateapj}
\usepackage{bm}
\usepackage{color}
\usepackage{graphicx}
\usepackage{amsmath}
\usepackage{tabularx} 
\begin{document}

\title{Electron Heating in a Relativistic, Weibel-Unstable Plasma}
\author{Rahul Kumar}
\author{David Eichler}
\author{Michael Gedalin}
\affiliation{Physics Department, Ben-Gurion University, Be'er-Sheba 84105, Israel}

\begin{abstract}
The dynamics of two initially unmagnetized relativistic counter-streaming homogeneous ion-electron plasma beams are simulated in two dimensions using the particle-in-cell (PIC) method. It is shown that current filaments, which form due to the Weibel instability, develop a large scale longitudinal electric field in the direction opposite to the current carried by the filaments as predicted by theory. Fast moving ions in the current filaments decelerate due to this longitudinal electric field. The same longitudinal electric field, which is partially inductive and partially electrostatic, is identified as the main source of acceleration of electrons in the current filaments. The transverse electric field,  though larger than the longitudinal one, is shown to play a smaller role in heating electrons, contrary to previous claims. It is found that, in 1D,  the electrons become strongly magnetized and are \textit{not} accelerated beyond their initial kinetic energy. Rather, the heating of the electrons is enhanced by the bending and break-up of the filaments, which releases electrons that would otherwise be trapped within a single filament and hence slow the development of the Weibel instability  (i.e. the magnetic field growth) via induction as per Lenz's law. In 2D simulations electrons are heated to about one quarter of the initial kinetic energy of ions. The magnetic energy at maximum is about 4 percent, decaying to less than 1 percent by the end of the simulation.  Most of the heating of electrons takes place while the longitudinal electric field is still growing while  only a small portion of the heating is a result of subsequent magnetic field decay.  The ions are found to gradually decelerate until the end of the simulation by which time they retain a residual anisotropy less than 10 percent.
\end{abstract}

\section{Introduction}
 Collisionless shocks forming in astrophysical environments are believed to be mediated by electromagnetic instabilities \citep{KennelSagdeev1967,Parker1961,Eichler1979,Blandford1987} in which ions are scattered by the resulting magnetic field fluctuations. In particlular, the Weibel instability \citep{Weibel}, which causes fast growth of strong magnetic field at small-scale in anisotropic plasma flow, has received much attention as the main isotropization mechanism that leads to shock transition in free-streaming ejecta from violent astrophysical events \citep{Medvedev1999,Achterberg2004,Yuri2006,Achterberg2007,Bret2009,Yalinewich,Rashid2012}.  As the dominant modes of the Weibel instability are less than the ion gyroradius, which renders them inefficient scatterers of ions, the longstanding question about its role in collisions shocks has been how well it competes with other mechanisms \citep[e.g.][]{Galeev1964,Blandford1987,Yuri2006}. Clearly mechanisms that require a preexisting magnetic field are questionable when the magnetic field is weak.  On the other hand, a slow shock mechanism may actually suppress a faster mechanism because it creates a broader shock transition, and thus has greater "reach" upstream of the shock. 
 
 The role of Weibel instability in forming the shock transition in weakly magnetized plasmas has recently been established through several numerical experiments \citep{Nishikawa2003,Fred2004,Nishikawa2005,spitkovsky2008,uri2009}. Numerical simulations of relativistic shocks have shown that the relatively less energetic upstream electrons are significantly heated to energy comparable to the energy of ions as they cross the foreshock, which appears to be in agreement with the electromagnetic observations of supernova remnants and gamma-ray bursts afterglows, where the high energy radiation is believed to be synchrotron radiation originating from gyration of high energy electrons in a magnetic fields significantly higher than the interstellar magnetic field \citep{pkumar2002,piran2005,Gehrels2012}. It however remains unclear how the upstream electrons are energized in the foreshock region, both in nature and in the numerical simulations of collisionless shocks, and whether the Weibel instability-induced magnetic field should persist over astrophysically significant length scales behind the shock.
 
   Numerical simulations suggest presence of large scale electric field in and around the current filaments forming just ahead of the shock and ions are found to decelerate due to this electric field. The same electric field that decelerates ions should also accelerate electrons \citep{Blandford1987,Yuri2006,Gedalin2008,Gedalin2012}. However, the details of the electric field and the acceleration mechanism have not been worked out. 

  In order to understand the heating of electrons in foreshock region of relativistic shocks we simulate, using the kinetic PIC method, development of the Weibel instability in two relativistic counter-streaming plasma beams, which resembles the precursor of a relativistic  shock, but simpler and more idealized and therefore more suitable for resolving certain fundamental questions. The beams are taken to be homogeneous and fully interpenetrating at the beginning, so the time development of the instability in our simulation imitates the spatial development of the instability in the foreshock regions, where later times in our simulation correspond the regions closer to the shock fronts. Our quantitative analysis suggests that the most of the heating of upstream electrons is due to the longitudinal electric field. The transverse electric field, though much stronger than the longitudinal, has negligible effect on net acceleration of electrons.  

    We have also inserted virtual test particles into  the code to act as numerical probes. They assist in the diagnostics of the mechanisms that are at work, and this will be discussed below.
 
\section{Numerical Simulation}
We use parallel version of the PIC code TRISTAN \citep{Buneman1993, Spitkovsky2005} to simulate  relativistic counter-streaming beams of ions and electrons in two dimensions. The simulation is initialized by placing ion-electron pairs at uniformly chosen random locations in a two dimensional square box in x-y plane. Initially ion and electron in each pair are moving in opposite directions along the x-axis with Lorentz factor $\gamma_0$. Half of the total number of pairs have ions moving along the positive x-axis while an equal number of another half pairs have ions moving along the negative x-direction, hence creating two oppositely streaming neutral plasma beams of equal intensity, where each beam has a net current.  The initial condition ensures that the simulation is initially charge and current free and that Maxwell equations are satisfied over scales where there are equal numbers of forward and backward moving electron-ion pairs, but at the grid scale, local fluctuations in the number of forward and of backward movers imply a net small scale current. This noise level, however, is reddened by smoothening the grid scale current. As the electrons and ions in the pairs separate from each other, the electric and magnetic field grow from the noise in the current generated by streaming electrons and ions. Since we simulate an initially unmagnetized plasma, only out-of-plane magnetic field $B_z$ and in-plane electric fields $E_x$ and $E_y$ are excited as the Weibel instability set in. We impose periodic boundary condition in both x and y directions for both particles and fields.  

Here we discuss results mainly from the two largest simulations we have attempted in terms of physical size and evolution time. Ion to electron mass ratio $m_i:m_e$ for the two reported simulations are 16:1 and 64:1, and are henceforth referred to as $M_{16}$ and  $M_{64}$, respectively. All the figures in this paper are for ion to electron mass ratio of 64:1, i.e. simulation $M_{64}$, unless otherwise stated. The physical sizes of the box for $M_{16}$ and $M_{64}$ are 500 c/$\omega_{pe}$x500 c/$\omega_{pe}$ and 1000 c/$\omega_{pe}$x1000 c/$\omega_{pe}$, respectively, where c is the speed of light in vacuum and $\omega_{pe}=\sqrt{4\pi n_0 e^2/m_e\gamma_0}$  is the initial electron plasma frequency, where $e$ is the charge of an electron and $n_0$ is the initial number density of electrons, or by charge neutrality, of ions. Both simulations were resolved to 1/10th of the initial electron skin-depths. Initially there are 32(8) particles per unit cell of the simulation box for  $M_{16}(M_{64})$ and the simulation was  evolved for 500 (1000) plasma time $\omega_{pe}^{-1}$. Initially the ions and electrons are moving along the x-axis with initial Lorentz factor $\gamma_0=10$ in the both cases. 

\section{Filamentation and electron heating}
At the very beginning  of the simulation the counter-streaming electrons, which are relatively lightweight, as compared to ions, and can relatively easily be deflected by the magnetic perturbations, are subject to the Weibel instability. Current due to the streaming electrons generate magnetic field and the electrons moving in the same direction are herded into the filaments by self created magnetic field. Growing current in the filaments induces an electric field opposite to the direction of current in accordance with Lenz's law. The induced electric field slows down the streaming electrons in the filaments and are scattered by the strong magnetic field around the filaments. During this stage relatively heavier ions continue streaming almost unaffected. As the electron Weibel instability stage ends in about 20 $\omega_{pe}^{-1}$, electrons are nearly thermalized to about their initial streaming kinetic energy $n_0m_e(\gamma_0-1)c^2$, creating a nearly isotropic electron background for still streaming ions. 

\begin{figure}
    \centering
    \includegraphics[width=0.5\textwidth]{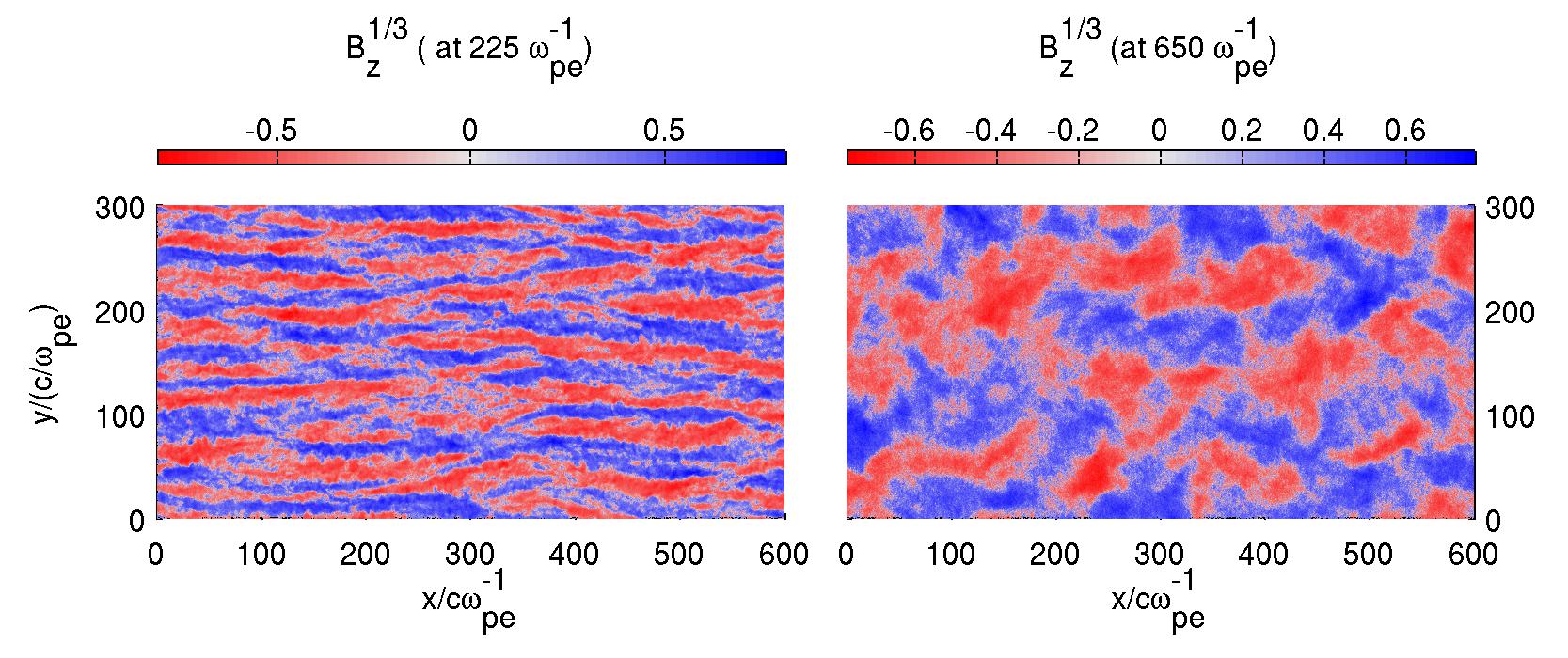}
    \caption{ Left panel: Structure of magnetic fields at 225 $\omega_{pe}^{-1}$. The magnetic field is still growing due to the Weibel instability and forms filamentary structures parallel to the streaming direction (referred to as filamentary phase). Most of the heating of electrons takes place during the filamentary phase. Right panel:  Structure of magnetic field at 650 $\omega_{pe}^{-1}$. Transverse size as well as the bending of filaments grow with time and eventually filaments get disoriented. Here, and henceforth, $B_z$ and all other electromagnetic fields (i.e., $E_x$ and $E_y$) are normalized to $(4\pi n_0 (\gamma_0-1)(m_i+m_e)c^2)^{1/2}$, unless otherwise specified.}
    \label{filaments}
\end{figure}

\begin{figure}
    \centering
    \includegraphics[width=0.5\textwidth]{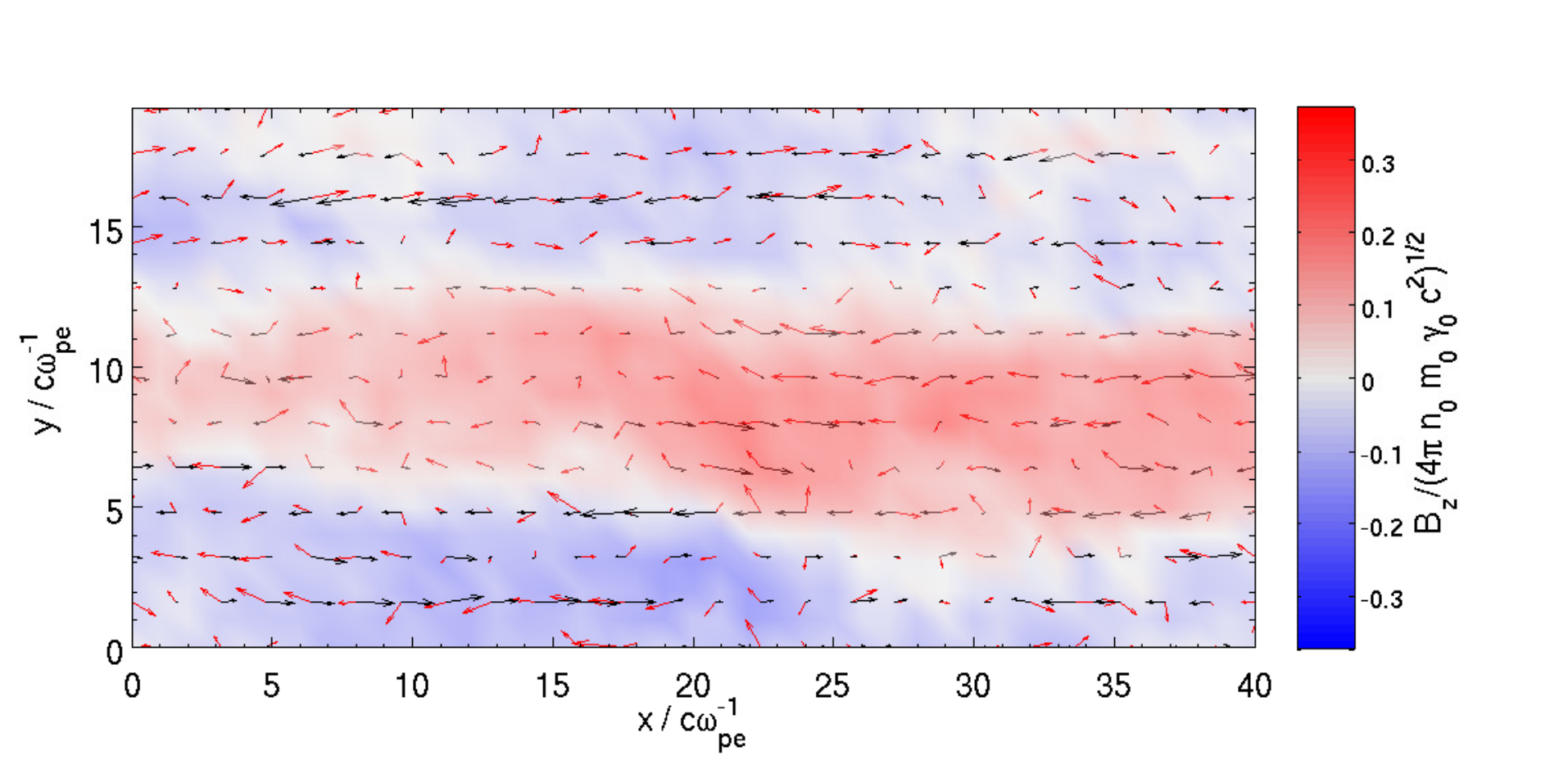}
    \caption{Local magnetic field, and the current due to electron and ions are shown in a spatial patch from the simulation $M_{64}$ ( $m_i:m_e$=64:1) at time 65 $\omega_{pe}^{-1}$. Direction of black and red arrows indicate direction of the local current due to the ions and electrons, respectively, at the tail of the arrows and length of the arrow is in proportion to the magnitude of the corresponding vector quantity. The local currents shown here are computed by taking the average velocity of all particles within a box of size $0.4 c\omega_{pe}^{-1}\times 0.4 c\omega_{pe}^{-1}$ centered at the tail of arrows. Inductive electric field opposes the current due to fast moving ions in the current filaments and is responsible for deceleration of ions as well as acceleration of electrons. }
    \label{JE}
\end{figure}

As the electron filamentation stage ends, relativistically counter-streaming ions undergo the Weibel instability and likewise current carrying ions are separated into long current carrying filaments along the initial streaming direction. The current filaments are non-stationary structures with slight bending in the transverse direction (figure \ref{filaments}), which move along with the current carrying ions with the strength of magnetic field being small near the filaments and large between two adjacent filaments. Again, the inductive electric field develops in the filaments opposite to the current due to the streaming ions. In addition to the inductive electric field along the filaments an electrostatic electric field develops around the filaments in the direction transverse to the streaming direction due to excess of positively charged ions in the filaments. The scale of this transverse electrostatic field is limited by the electrons that cloud around the filaments and remains $\sim c/\omega_{pe}$, which grows with time as the electrons are accelerated to higher energies. In Figure \ref{JE} we show the details of the local magnetic field and the local current due to ions and electrons in a small spatial part of the simulation box. The apparent correlation between the motion of ions and electrons (electrons move opposite to their current) suggests that the electrons are accelerated in the direction of streaming ions. 

The spatial distribution of electrons closely follows the distribution of ions. Charge neutralizing electrons clump in rather positively charged filaments and stream along with the ions. The electrons in the filaments move along the inductive electric field created by the streaming ions and consequently get energized. Ions on the other hand move against the electric field induced by themselves and hence lose their energy. Since the current filaments are non-stationary and turbulent at very small scale, and so are the field structures, the local details of energy gain and loss can deviate significantly from the one mentioned above and hence an statistical approach is needed to understand the systematic acceleration of electrons. We therefore compute spatial average of various quantities and correlates to quantify the presence of systemic heating of electrons at scales much larger than their skin depth.  

\begin{figure}
    \centering
    \includegraphics[width=0.5\textwidth]{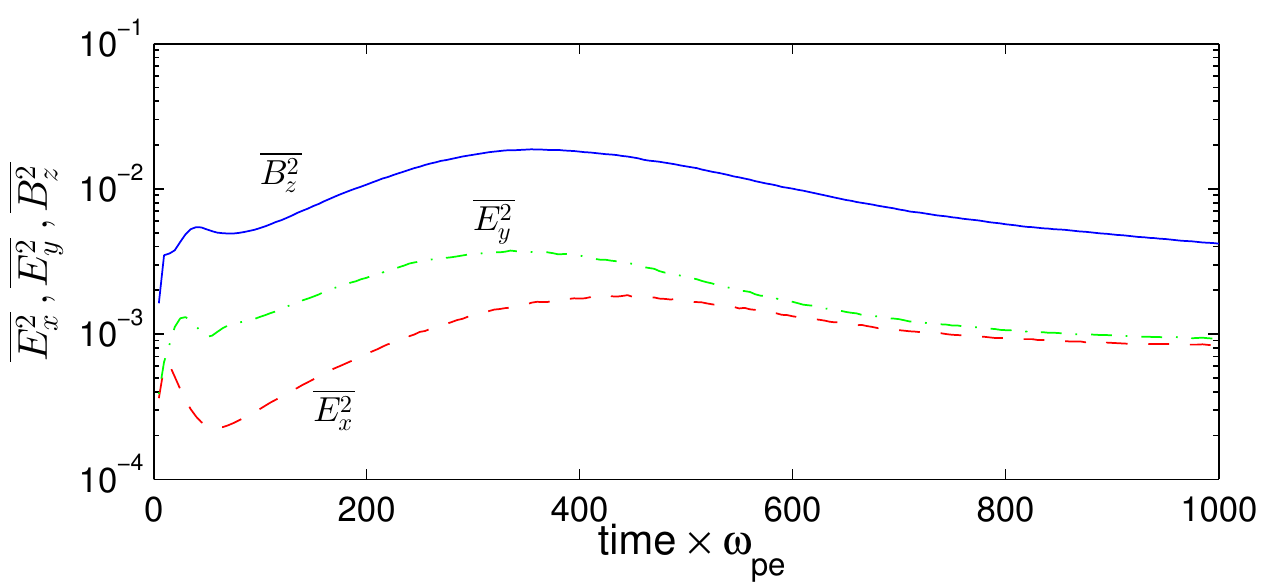}
    \caption{Temporal variation of the mean square (averaged over the physical domain of the simulation) of the electromagnetic field components $E_x$ (dashed red), $E_y$ (dotted dashed green), and $B_z$ (solid blue) are normalized to $4 \pi n_0(m_i+m_e)(\gamma_0-1)c^2$. Amongst the electromagnetic fields, the magnetic field gains most of the kinetic energy lost by the ions to the fields, which holds positively charges ions in the filaments together against the electrostatic electric field, followed by mostly electrostatic transverse electric field $E_y$ and longitudinal electric field $E_x$.}
    \label{fld_mag}
\end{figure}

During the linear stage of the instability not all the energy that is lost by ions goes into heating electrons. A significant part of the total kinetic energy goes into the electromagnetic fields which are responsible for mediating collective interactions between the charged particles. In figure \ref{fld_mag} we show how much of initial kinetic energy goes into different components of the electromagnetic fields. During the ion filamentation stage, which is mainly magnetic in nature, energy in the magnetic fields reaches few percent of the total initial kinetic energy. Longitudinal electric field $E_x$ is the weakest component, but as shown later, is the most important for the net transfer of kinetic energy from ion to electrons. The longitudinal electric field $E_x$ has a curl in the direction of magnetic field which predominantly contributes to the growth of the magnetic field. In other words, the inductive electric field $E_x$ accompanies the growth of the magnetic field and accelerates electrons hence leaving it the weakest amongst all growing electromagnetic field components. As the non-linearity sets in and the filaments are bent and broken, $E_x$ and $E_y$ both converge to a similar value since disruption and disorientation of current filaments wipe out any preference in the direction of the electric field in the x-y plane (figure \ref{filaments}). 

\subsection{electron heating estimate} 
The inductive longitudinal electric field in the direction opposite to the current in plasma features in the linear theory of the Weibel instability itself. In fact it is the curl of this longitudinal electric field which accompanies the growth of magnetic field in accordance with the Faraday's law of induction. Presence of a large scale electric field in a conducting plasma naturally predicts transfer of kinetic energy from ion to less energetic electrons. Here we estimate the net energy gained by electrons due to the inductive electric field in a one dimensional linear theory of the Weibel instability. In the two dimensional case simulated here electromagnetic field fluctuations due to the electrostatic wave modes parallel to the x-axis (streaming direction) and the oblique modes also become comparable to the transverse modes, and eventually lead to the disruption of current filaments \citep{Rashid2012}. Here we consider the heating solely due to the transverse modes with the wave vector $\vec{k}$ along the y-axis and electric field $\vec{E}$ along the x-axis. We assume homogeneity along the x-axis, that is averaging out the fluctuations along the x-axis, hence reducing the problem to one spatial dimension. 

Let $p_x(y)$ be the x-momentum of an electron at any given location y. The electric field $E_x(y)$ at any given y determines the rate of change of x-momentum of an individual electron located at y, i.e. $dp_x(y)/dt=eE_x(y)$. Averaging the x-momentum of all electrons at a certain y gives, 

\begin{equation}
d \bar{p}_x(y) /dt=eE_x(y)
\label{motion}
\end{equation}
where $\bar{p}_x(y)=\int p_x f_e(y,\vec{p},t)d^3p/\int f_e(y,\vec{p},t) d^3p$ with $f_e$ being the distribution function of electrons which is assumed to depend on y only. The electromagnetic fields $E_x$ and $B_z$, as well as the mean x-momentum of electrons $\bar{p}_x$, can all be written as sum of sinusoidal modes of various wavelengths, but with $B_z$ quarter a wavelength phase shifted from $E_x$ and $\bar{p}_x$, since $B_z$ is maximum between two adjacent current filaments but $E_x$ and $\bar{p}_x$  peak in the current filaments. We first consider a single sinusoidal mode of wave number k and write $B_z=B^k \sin (ky)$, where $B^k$ is time dependent amplitude of the magnetic field $B_z$ in the mode of wave number k. From Maxwell's equations we relate the electric and magnetic field as $ kcE_x =\cos (ky)\partial B^k / \partial t$. Decomposing the linear equation \ref{motion} into Fourier components and then substituting for the electric field we find    
\begin{equation}
 \partial \bar{p}^k_x / \partial t=(e/kc) \partial B^k / \partial t
\label{pamp}
\end{equation}
 where $\bar{p}_x^k$ is the amplitude of average x-momentum $\bar{p}_x$ in the mode of wave number k. Equation \ref{pamp} enables us to express instantaneous amplitude of average three-momentum fluctuation in terms of the amplitude of magnetic fluctuation as 
 \begin{equation}
  \bar{p}^k_x =eB^k / kc +p_{x0}
  \label{pB}
  \end{equation}
  
  Equation \ref{pB} is of  central importance: It implies that the gyroradius of an electron, in the limit that
$p_x>>p_{xo}$, is roughly the wavelength times $1/2\pi$. As we shall see below, this is confirmed by the simulations. The implication is that the field growth is \textit{not stopped} by the magnetization of the electrons, as conjectured by Lyubarsky and Eichler ( 2006), because the electrons are at all times only marginally magnetized.  That the energy of the electrons keeps pace with the field growth, so that they never get highly magnetized, is the 
essential reason the electrons get heated as much as they do.

It is also shown that the marginal magnetization on the other hand, is enough to suppress the electron heating in one dimension, because the filaments remain exactly straight, and the electrons are accelerated exactly along the filaments.  By contrast, in more than one dimension, the existence of oblique modes allows the filaments to bend, and it becomes harder for an electron to remain within a single filament. The countercurrents of the electrons are then less likely to cancel those of the ions and the field growth can proceed to larger length scales.
  
\begin{figure}
    \centering
    \includegraphics[width=0.5\textwidth]{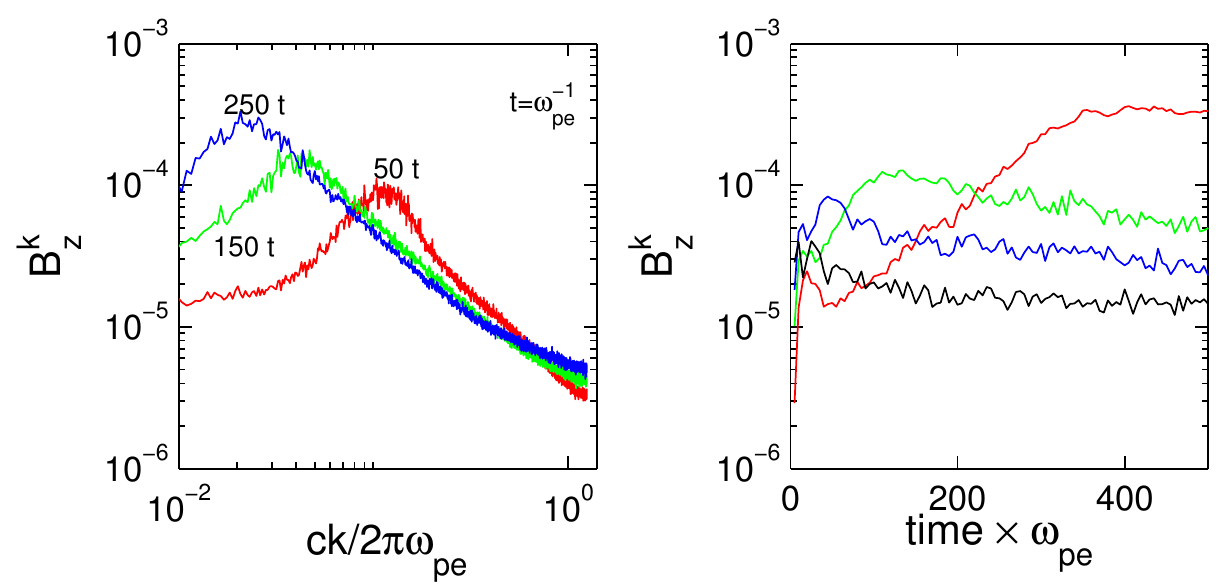}
    \caption{ Left Panel: average $B_z^{k}$ (amplitude of magnetic field in mode of wavenumber k) as a function of wave number k at time 50, 150, and 250 $\omega_{pe}^{-1}$ are shown by red, blue, and green curves, respectively.  The spectrum is obtained by first computing one dimensional discrete Fourier transform of magnetic field in several slices of the simulations box along the transverse direction and then taking average of the magnetic field amplitudes in any given mode k over all these slices. Right panel: evolution of average amplitude of transverse magnetic fluctuation $B_z^{k}$ for $2\pi/k= 10, 2, 1, 0.5 c\omega_{pe}^{-1}$ are shown by red, green, blue, and black curves, respectively. The magnetic field in any given mode grows exponentially and then slowly decays. }
    \label{Bkvstime}
\end{figure}

\begin{figure}
    \centering
    \includegraphics[width=0.5\textwidth]{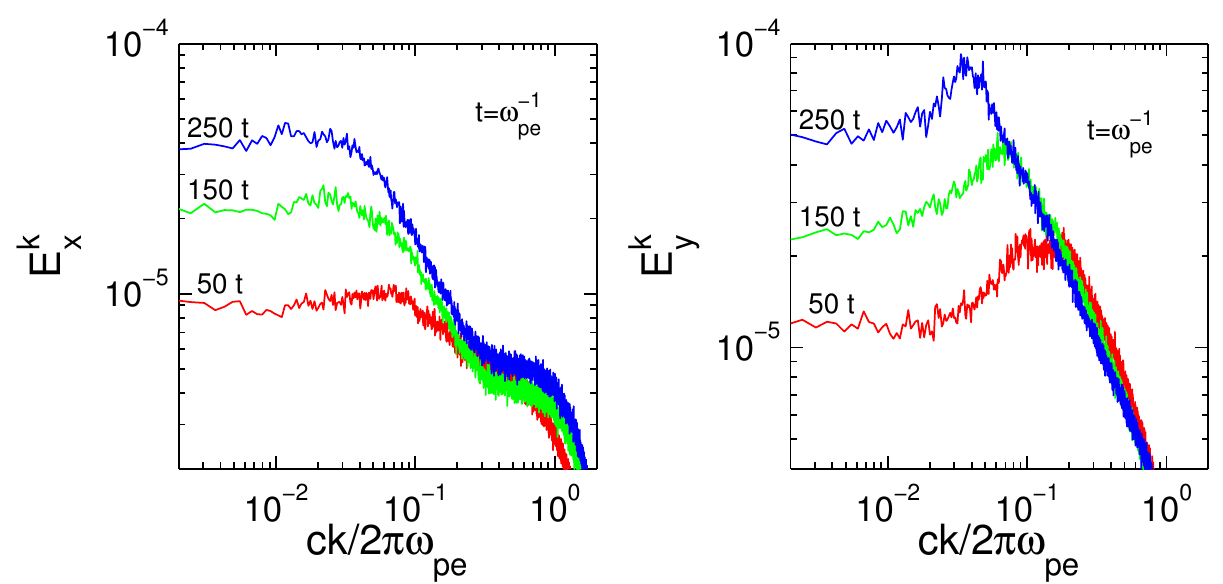}
    \caption{Transverse spectrum, as in the left panel of figure \ref{Bkvstime}, of longitudinal electric field $E_x$ and transverse electric field $E_y$ as a function of wave number k at time 50, 150, and 250 $\omega_{pe}^{-1}$ are shown by red, blue, and green curves, respectively.}
    \label{specExEy}
\end{figure}

   At any instant, the rate of change of average kinetic energy of an electron $U_e=m_e(\gamma_e-1)c^2$ due to the longitudinal electric field $E_x$ is given by the spatial average (y-average) of $eE_x\bar{v}_x$. For a single mode, defining $U^k_e$ as y-average of $c\bar{p}_x^k$ we obtain (after squaring the equation \ref{pB} and then taking the derivative with respect to time)
 \begin{equation}
 U^k_edU^k_e/dt \approx [e^2c^2/(kc)^2] B^k \partial B^k / \partial t 
 \label{rate}
 \end{equation}
 The linear theory of Weibel instability predicts exponential growth of the amplitude of magnetic field fluctuation $B^k$ at all length scales. However, the maximum strength of magnetic field at any given length scale is constrained by the available current in the plasma at that length scale. Consequently, the growth of magnetic field amplitude $B^k$ at a length scale $1/k$ reaches a saturation amplitude $B^k_{peak}$, which is inversely proportional to the wave number $k$\footnote{the maximum achievable net current in the filaments of transverse size 1/k is limited by the density of streaming ions in that length scale, which is $\sim n_0ec/k$, implying that $B^k_{peak} \propto 1/k $ \citep{Kato2005,Gedalin2010} }. Since the amplitude of magnetic field $B^k$ at smaller scales saturates first, while the magnetic field at larger scales are still growing, the net magnetic field at any instant is dominated by the magnetic field in the largest transverse scale achieved during the Weibel instability by that time. In Figure \ref{Bkvstime} we show the time evolution of transverse spectrum of the magnetic field structure in the simulation which confirms the exponential growth and saturation of the magnetic field at length scales smaller than a few times the skin depth of ions. In figure \ref{specExEy} we show the transverse spectrum of $E_x$ and $E_y$ as well. It shows the exponential growth and saturation of the transverse electric field, and that the transverse electric field peaks at half the wavelength magnetic field peaks, since the transverse electric field is symmetric with respect to the current filaments. The transverse spectrum of longitudinal electric field in figure \ref{specExEy} does not show saturation because of growth of electrostatic parallel modes, as shown later, which contributes to further growth of the longitudinal electric field  .
 
  The heating due to the longitudinal electric field $E_x$ is significant only until the filaments prevail. As the transverse scale of the filaments reach ions skin-depth, bending of the filaments in the transverse direction becomes significant. The current filaments start to break, get randomly oriented in x-y plane, and the rate of heating of electrons is substantially reduced. In order to obtain an estimate for the total energy acquired by the electrons by the end of the filamentation it suffices to integrate equation \ref{rate} with respect to time for the smallest wavenumber $k_{min}$ achieved during the Weibel instability with the electron heating in progress, and integrating it up to the time saturation is achieved in this mode, that is to say until the heating due to this mode is significant. Following these prescriptions for the net heating estimate we get,
 \begin{equation}
{U_e^{f}}^2 \approx m_i \gamma_0c^2 (\omega_{pi}/k_{min}c)^2 ({B_{peak}^{k_{min}}}^2/ 8 \pi n_0)
\label{uef}
\end{equation}
 where $U_e^f$ is the average energy of an electron towards the end of the heating process during the linear phase. 
 
\begin{figure}
    \centering
    \includegraphics[width=0.5\textwidth]{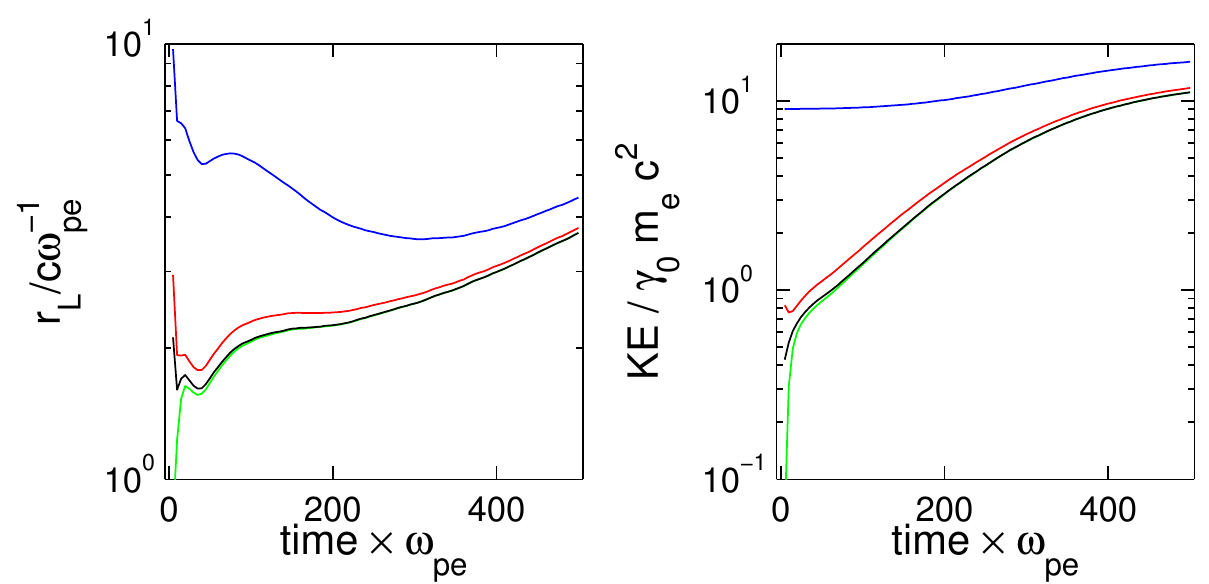}
    \caption{Left Panel : The mean Larmor radii of electrons as well as test particles (charged particles that do not participate in the plasma dynamics but respond to the local electromagnetic field) with charge to mass ratio 0.1 than that of the electrons are shown in units of instantaneous skin depth of electrons. The red curve shows Larmor radius to electron skin depth ratio for electrons, while black, blue, and green curves show the same for test particles with initial kinetic energy 0.01, 0.1, and 10 times that of the electrons. During the linear stage of the instability Larmor radius of electrons as well as the test particles gaining energy from the longitudinal electric field remains about the instantaneous skin depth of the electrons. Right Panel: Temporal variation of mean kinetic energy of the electrons and test particles in units of $\gamma_0 m_e c^2$. Color code for the curves is same as in the left panel. Energetic particles that have Lamor radii much larger than the spacing between the filaments, such as the test particles with initial kinetic energy ~$(\gamma_0-1)m_ec^2$ (shown in blue) do not get efficiently energized by the longitudinal electric field. }
    \label{rlsd}
\end{figure}

 As compared to the gyro-frequency of electrons, the transverse scale of filaments grow relatively slowly with time. Electrons gain energy in the current filaments and their Larmor radii get enlarged. However, at any instant during the linear stage of the Weibel instability the enlarged Larmor radii of electrons due to inductive electric field in the current filaments cannot substantially exceed the transverse size of the current filaments. This limitation on electron Larmor radii is due to the fact that for an electron that has Larmor radius much larger than the spacing between two adjacent filaments, energy gained in one filament is lost in the neighboring filaments which carries opposite current and hence $E_x$ directed in opposite sense. Additionally, as suggested by equation \ref{pB}, the energy imparted by the inductive longitudinal electric field in the largest length scale is just enough to keep the Larmor radii of electrons at any instant about $1/2\pi$ times the transverse size of the current filaments. As the electrons get accelerated to higher energies their skin-depth also increases in the same proportion. Indeed, as observed in the simulations (figure \ref{rlsd}), the ratio of the Larmor radius to the instantaneous skin depth of the electrons, which is also about the $1/2\pi$ times the transverse size of the current filaments, is nearly constant during the linear stage of the Weibel instability. It suggests that during the filamentation stage the inductive heating of electrons proceed such that $n_0U_e \gtrsim B^2/8\pi$ remains satisfied. 
   
 The growth of magnetic field and transverse size of the filaments continue until the transverse scale of the filaments becomes comparable to the ions skin depth $c/\omega_{pi}$ after which the filaments are disoriented and the heating of electrons due to longitudinal inductive field is substantially reduced (in the two dimensional case electrons continue to gain energy due to decay of magnetic field, but at a much reduced rate). It suggests that for the purpose of estimating the total heating due to inductive electric field we can take $ck_{min} \sim \omega_{pi}$. This along with the condition that during the heating $nU_e \gtrsim B^2/8\pi$  suggest that $U_e^f \lesssim m_i\gamma_0 c^2$ (from equation \ref{uef}), that is to say the electrons are accelerated to energy comparable to the energy of ions during the linear stage of the Weibel instability due to the inductive longitudinal electric field. In the following we present results from the PIC simulation which confirms that the heating of electrons due to the inductive longitudinal field is indeed significant, though there is additional heating due to the electrostatic modes which appear in the two dimensional analysis and is essential to the heating of electrons.

\subsection{Net work done by the electric field} 
\begin{figure}
    \centering
    \includegraphics[width=0.5\textwidth]{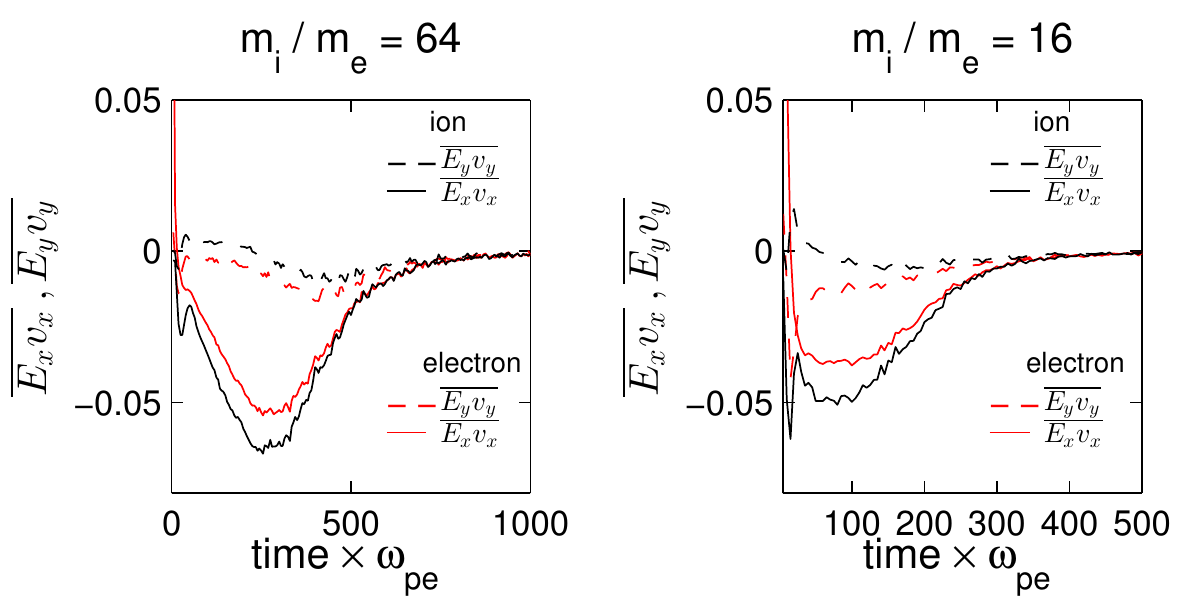}
    \caption{The rate of energy gained (lost) by electrons (protons), as a function of time, due to $E_x$ and $E_y$, i.e. $\overline{E_xv_x}$ and $\overline{E_y v_y}$, respectively (in units of $c^2\sqrt{4\pi \gamma_0 m_e}$),  are shown by solid and dashed red (black) curves, respectively . The quantities $\overline{E_xv_x}$ and $\overline{E_y v_y}$ are the mean values of $E_xv_x$ and $E_yv_y$, respectively, computed for the sample of particles (few percent of the total number of particles) which were initially homogeneously distributed over the simulation box. The left and right panels are for ion to electron mass ratios 16:1 and 64:1, respectively. Though the transverse electric field $E_y$ is much larger in magnitude than the longitudinal electric field $E_x$ ( figure \ref{fld_mag}), its contribution to the electron acceleration and ion deceleration (dashed lines) is negligible as comparable to the heating due to $E_x$.}
    \label{edotv}
\end{figure}

While the magnetic field scatter the charged particles by altering their trajectories, the only source of acceleration of the plasma particles is electric field in the lab frame. From our PIC simulations we separate out work done by the two orthogonal components of the electric field, namely $E_x$ and $E_y$, on electrons as well as on protons.  As evident from Figure \ref{edotv},  in both the simulations reported here the change in total kinetic energy of electrons and protons is apparent to be mainly due to the longitudinal electric field $E_x$. The net work done by the transverse electric field $E_y$, which is rather stronger than the longitudinal electric field (Figure \ref{fld_mag}), is vanishingly small \footnote{The implication is that the net energization of electrons due to electrons falling into the growing and merging filaments, as suggested by \cite{Hededal2004} and \cite{Anatoly2008a}, is rather small during the filamentation stage of the Weibel instability. }.

\begin{figure}
    \centering
    \includegraphics[width=0.5\textwidth]{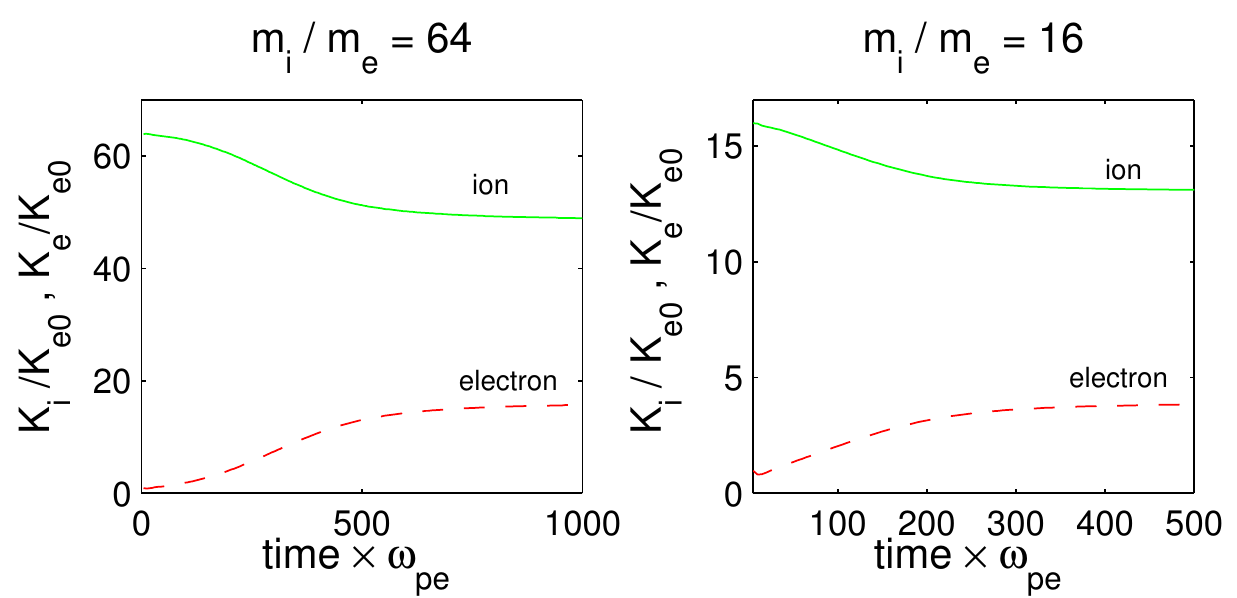}
    \caption{Evolution of mean kinetic energy of electrons $K_e$ (dashed red ) and ions $K_i$ (solid green) are shown for two different ion to electron mass ratios (normalized to the initial kinetic energy electrons $K_{e0}$) . Electrons heating due to the Lenz electric field is significant until the ion current filaments are disrupted. By the end of the simulation electrons are heated to about 1/3 of the ions kinetic energy in both cases consider here.  }
    \label{KE}
\end{figure}

In figure \ref{KE} we show the time evolution of the total kinetic energy of electrons and ions. It shows that most of the energy exchange between ions and electrons takes place during the filamentation stage of the instability, when magnetic field is still growing (figure \ref{fld_mag}) in strength. We have run the simulation for long enough to capture most of the heating phase and long after the rapid heating phase ends, electrons are found to have acquired a substantial part of the ions's kinetic energy, in agreement with the estimate presented above. Even long after the breaking of filaments, ions are not completely thermalized and continue to lose energy and the heating of electrons continues, though at a much lower rate. 

Bending of the current filaments is apparent from the very beginning of the Weibel instability. The bending of current filaments results in mixing of the longitudinal and transverse components of the electric field. Two dimensional linear theory of the Weibel instability predicts the growth of waves with wave vector at oblique angle with respect to the streaming direction \citep{Rashid2012}, and we suggest that this can account for the bending of the filaments. The filaments are also susceptible to the Buneman instability, which leads to the growth of resonant waves parallel to the streaming direction. The oblique and parallel modes are rather electrostatic in nature which implies that the the true nature of the longitudinal and transverse electric field (i.e., electrostatic or induced) is rather mixed and are due to several waves growing at the same time, though at different rates (see the section \ref{pllmode} ). In order to quantify the role of different types of growing waves in the relativistic counter-steaming plasma in heating of electrons we separate (Helmholtz decomposition) the electric field into the rotational and compressive part, i.e. $\vec{E}=\vec{E}^c+\vec{E}^r$, such that $\nabla \cdot \vec{E}^r=0$ and $\nabla \times \vec{E}^c=0$. If the only waves growing in the counter-streaming plasma were the purely transverse Weibel modes then the longitudinal electric field would be purely inductive in nature and could solely be described by a rotational electric field. However, in case of oblique and parallel electrostatic waves appearing in the plasma along with the purely transverse waves, the longitudinal electric field can not completely be described by a divergence free electric field.  

\begin{figure}
    \centering
    \includegraphics[width=0.5\textwidth]{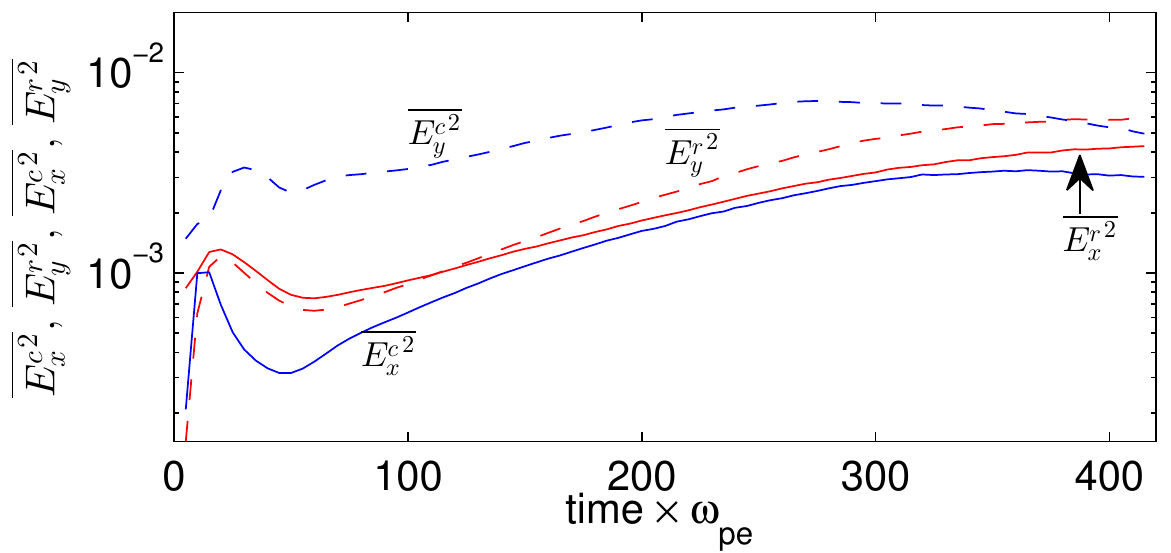}
    \caption{Temporal variation of the mean square (as in the figure \ref{fld_mag}) of the electric field components $E_x^r$, $E_x^c$, $E_y^r$, and $E_y^c$ ($\vec{E}=\vec{E}^r+\vec{E}^c$, such that $\nabla \cdot \vec{E}^r=0$, and $\nabla \times \vec{E}^c=0$. Subscript x and y indicate components along longitudinal (streaming direction) and transverse direction, respectively) are shown by the solid red, solid blue, dashed red, and dashed blue curves, respectively. During the filamentation stage the transverse electric field $E_y$ is mostly electrostatic and the longitudinal electric field $E_x$ is partially inductive and partially electrostatic.}
    \label{rotpotE2}
\end{figure}

\begin{figure}
    \centering
    \includegraphics[width=0.5\textwidth]{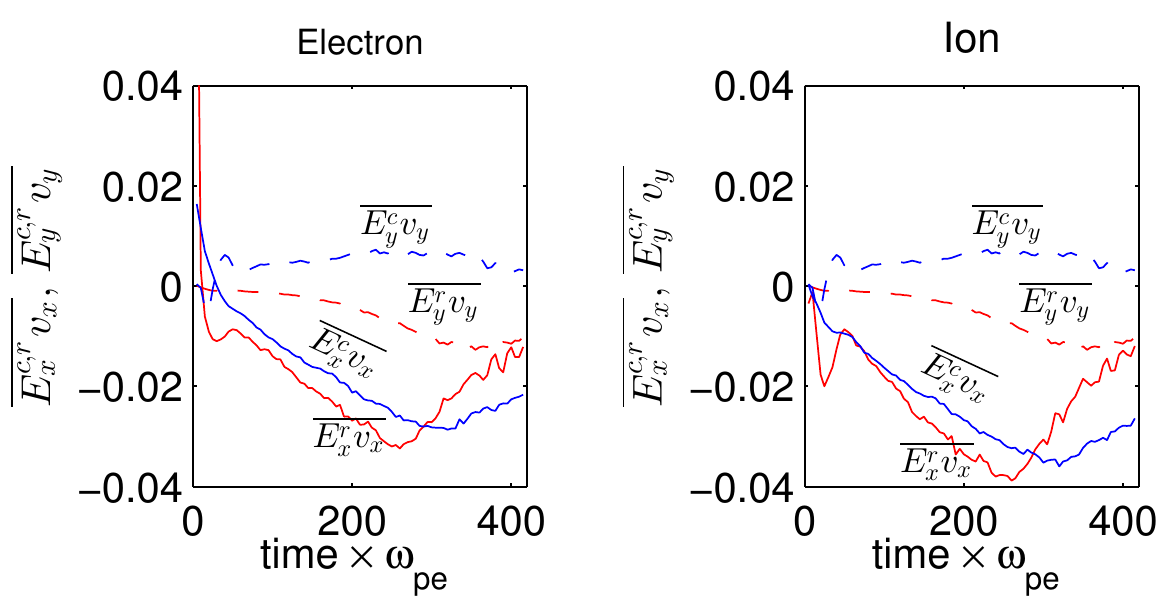}
    \caption{The left (right) panel shows the instantaneous rate of energy gained (lost) by the electrons (protons) due to the rotational and compressive components of the longitudinal electric field in solid red and solid blue, respectively. The dashed red and blue curves shows the same due to the rotational and compressive components of the transverse electric field $E_y$, respectively. Normalization and computation of the mean shown here are same as in the figure \ref{edotv}.}
    \label{rotpot}
\end{figure}

 In figure \ref{rotpotE2} we show to what extent the longitudinal and transverse components of the electric field is electrostatic and inductive in nature. It is evident from the figure that during the linear stage of the instability the transverse electric field is mostly electrostatic is nature and is due to excess of positive charges in the current filaments. The longitudinal electric field on the other hand is partially inductive and partially electrostatic, but more of an inductive in nature. In figure \ref{rotpot} we show the work done by the rotational and compressive part of each components of the electric field. We find that the work done by the compressive and rotational part of the longitudinal electric field are comparable, suggesting that the role of resonant heating by the electrostatic waves is comparable with the inductive heating due to  transverse Weibel modes. 

\subsection{Parallel and oblique modes}
\label{pllmode}
\begin{figure}
    \centering
    \includegraphics[width=0.5\textwidth]{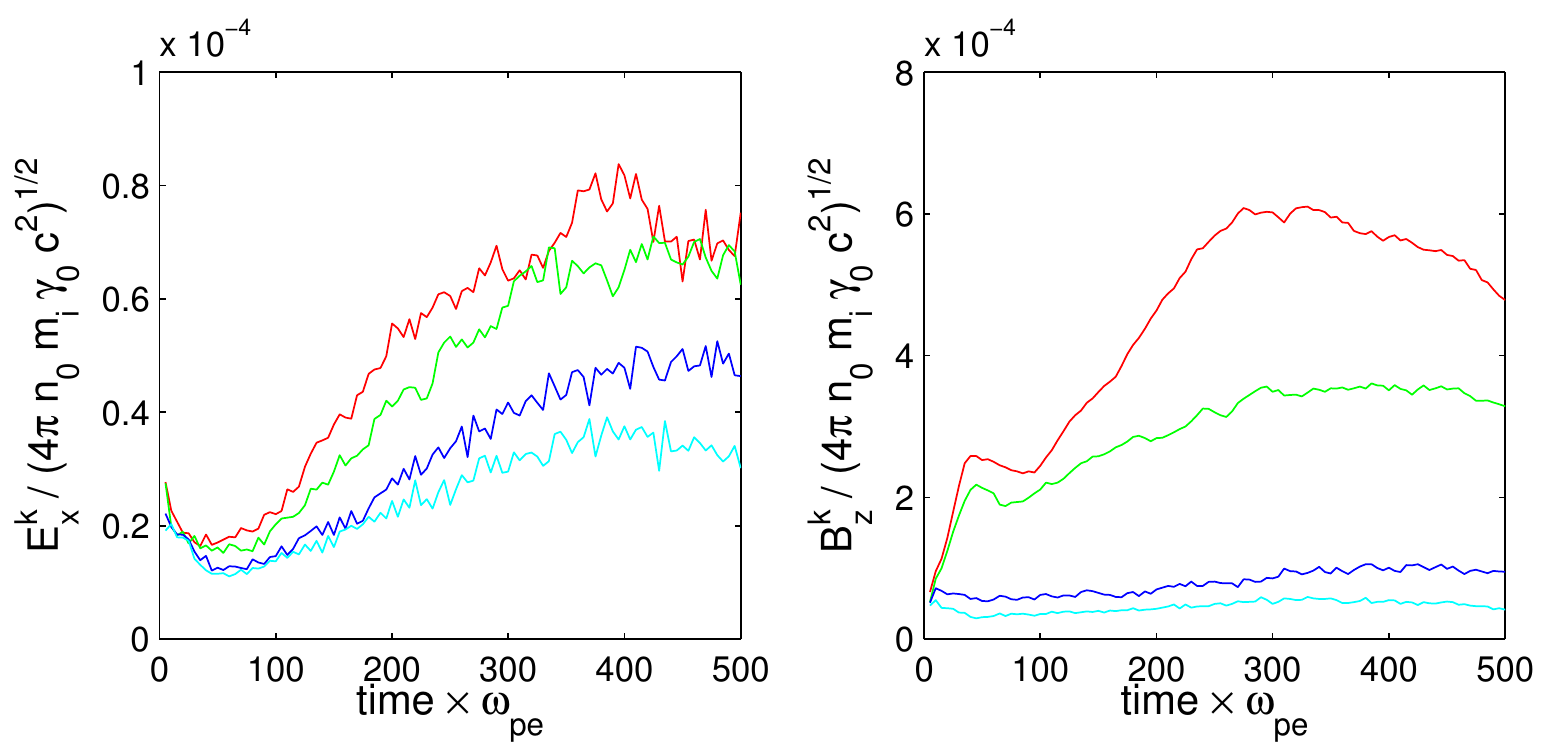}
    \caption{The left (right) panel shows temporal evolution of the average amplitude of sinusoidal variation in the longitudinal electric (magnetic) field $E_x^k$ ($B_z^{k}$) with the wave vector being along the streaming direction (x-axis). The red, green, blue, and cyan colored curve show the field amplitude of modes with $2\pi/k$= 410, 90, 16, and 8 $c/\omega_{pe}$, respectively. The average amplitude shown here are obtained after taking average of amplitude of any given mode over an ensemble of several slices parallel to the y-axis at equally spaced locations along the x-axis.  }
\label{parallel_mode}
\end{figure}

We perform two dimensional spectral analysis of the electromagnetic field components for the simulation $M_{64}$ to show the simultaneous growth of several other waves in addition to the purely transverse Weibel waves. In figure \ref{parallel_mode}, we show the average amplitude of sinusoidal variation in electromagnetic fields along the streaming direction. The amplitude of variation in the magnetic field serves as a proxy for bending of the current filaments since the parallel modes are mostly electrostatic and do not contribute to the growth of the magnetic field. As evident from the figure \ref{parallel_mode}, the growth of electric field in the same length scale is faster than the growth of the magnetic field, hence by comparison a rather stronger electric field in waves with their wave vector along the longitudinal direction can be attributed to the growth of electrostatic modes along the streaming direction in addition to the contribution from the bending of current filaments.

\begin{figure}
    \centering
    \includegraphics[width=0.5\textwidth]{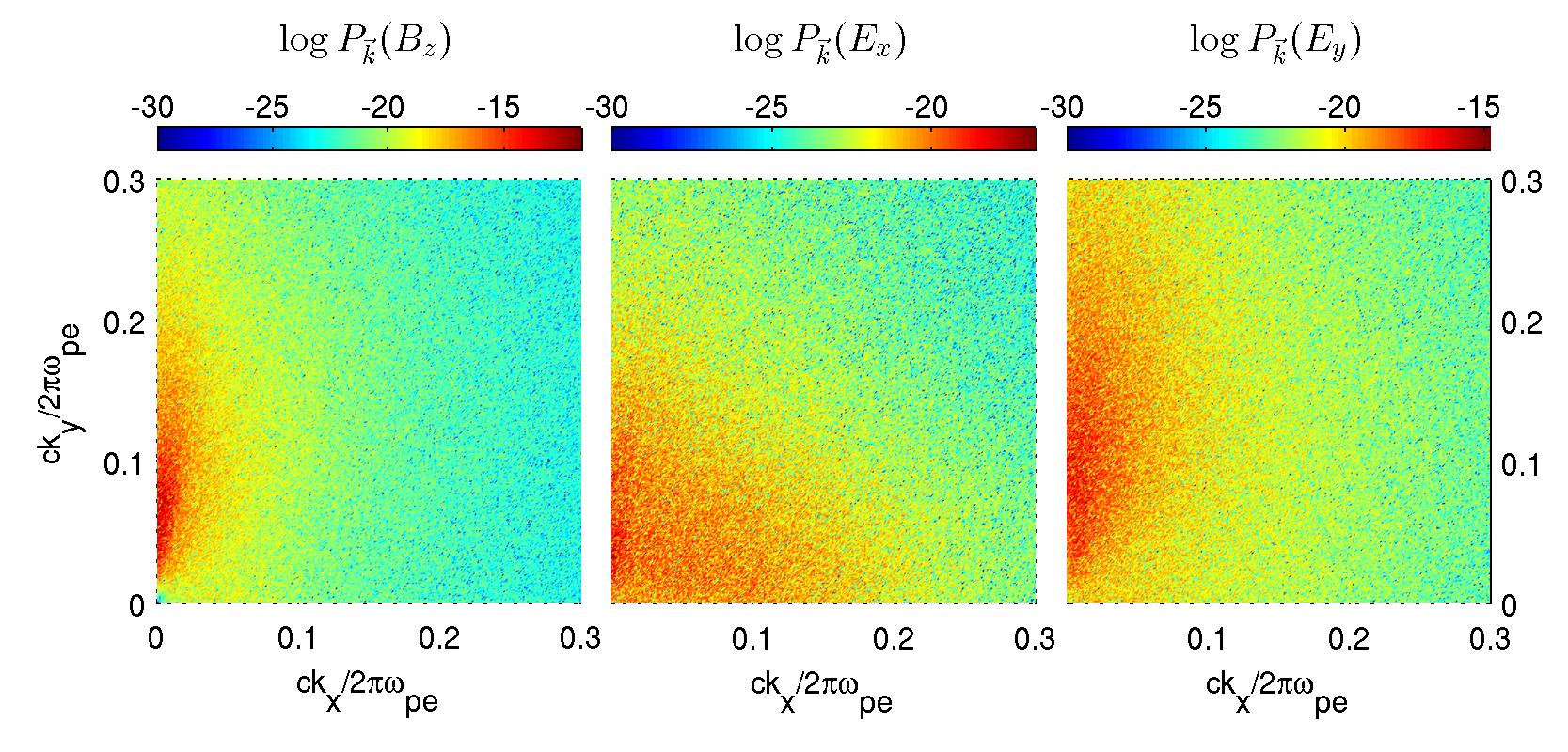}
    \caption{ Power spectrum for $B_z$, $E_x$, and $E_y$ at 100 $\omega_{pe}^{-1}$ is shown in the left, middle, and right panel panel, respectively. The power spectrum shows broadband nature of the Weibel instability. The spectrum of $B_z$ shows that the transverse waves are mostly magnetic in nature. The existence of parallel and oblique modes, which are rather electrostatic in nature, is apparant from the spectrum of $E_x$.}
\label{power_spectrum}
\end{figure}

\begin{figure}
    \centering
    \includegraphics[width=0.5\textwidth]{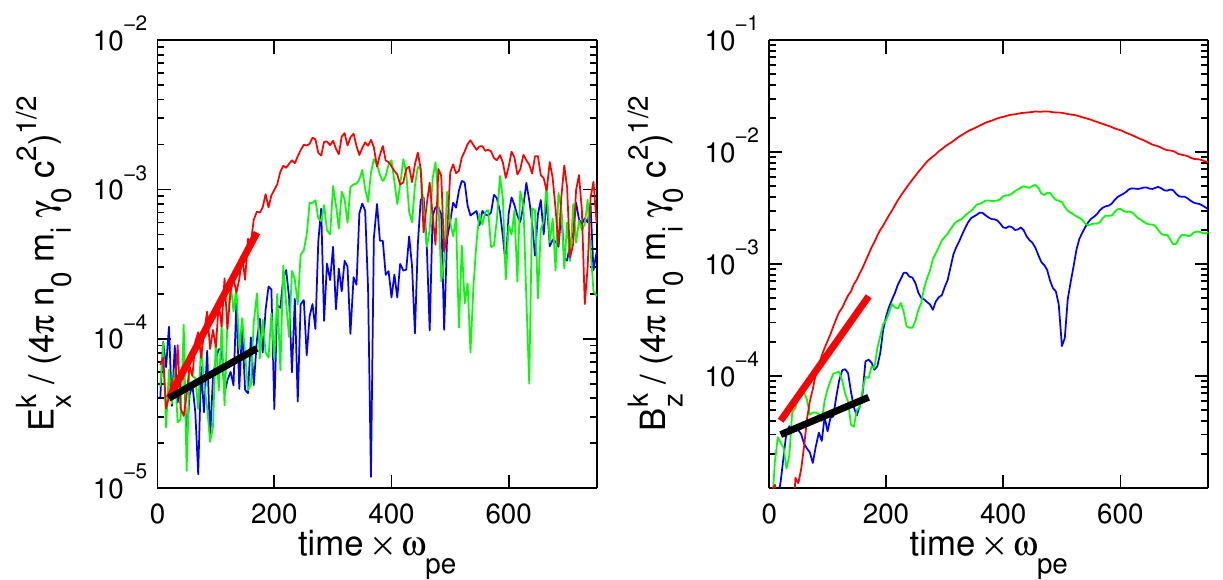}
       \caption{The left (right) panel shows temporal evolution of the amplitude of sinusoidal variation (obtained from two dimensional Fourier spectrum of electromagnetic fields) in the longitudinal electric (magnetic) field $E_x^k$ ($B_z^{k}$) with wave vector being along x-axis, y-axis, and at an equal angle with respect to the x and y axes . Specifically, the red, green, and blue curves correspond to $(2\pi/k_x, 2\pi/k_y) = (0, 100), (100, 100), and (100, 0) c/\omega_{pe}$, respectively. The thick solid line segments in red and blue indicate the growth rate predicted by the linear theory (taken from \cite{Rashid2012} for electron temperature $m_e/m_i$ times the ion temperature) for the transverse and the oblique wave, respectively. }
\label{oblique}
   \end{figure}

In figure \ref{oblique} we show the growth of magnetic and electric field in purely transverse, purely longitudinal, and an oblique mode. The general trend is that the purely transverse modes grow the fastest and the rate of growth decreases with the angle between the wave vector and y-axis. The oblique modes, like the purely transverse mode, are suppressed in the linear regime by the inductive response of the electrons which are accelerated by, and therefore move along the inductive electric fields.  Note, however, that the purely transverse mode stops growing and at time of about 250 $\omega_{pe}$ (figure) while the oblique mode continues to grow and  in fact catches up with the purely transverse mode.  This can be interpreted as follows: The transverse mode, being the fastest growing, is the first to trap the electrons into the resonance, so that the number of resonant electrons actually increases.  This includes those electrons that resonated with the oblique modes until  trapped by the  purely transverse mode. We can say that these electrons are in a non-linear resonance with the purely trapped mode.  At this point, the growth of the purely transverse mode is impeded by the even larger number of non-linearly resonant electrons, so it stops growing.The oblique modes, in contrast, suffer less from the inductive effects of the electrons so they keep growing.  until they have spoiled the alignment of the filaments made by the purely transverse mode.

\begin{figure}
    \centering
    \includegraphics[width=0.5\textwidth]{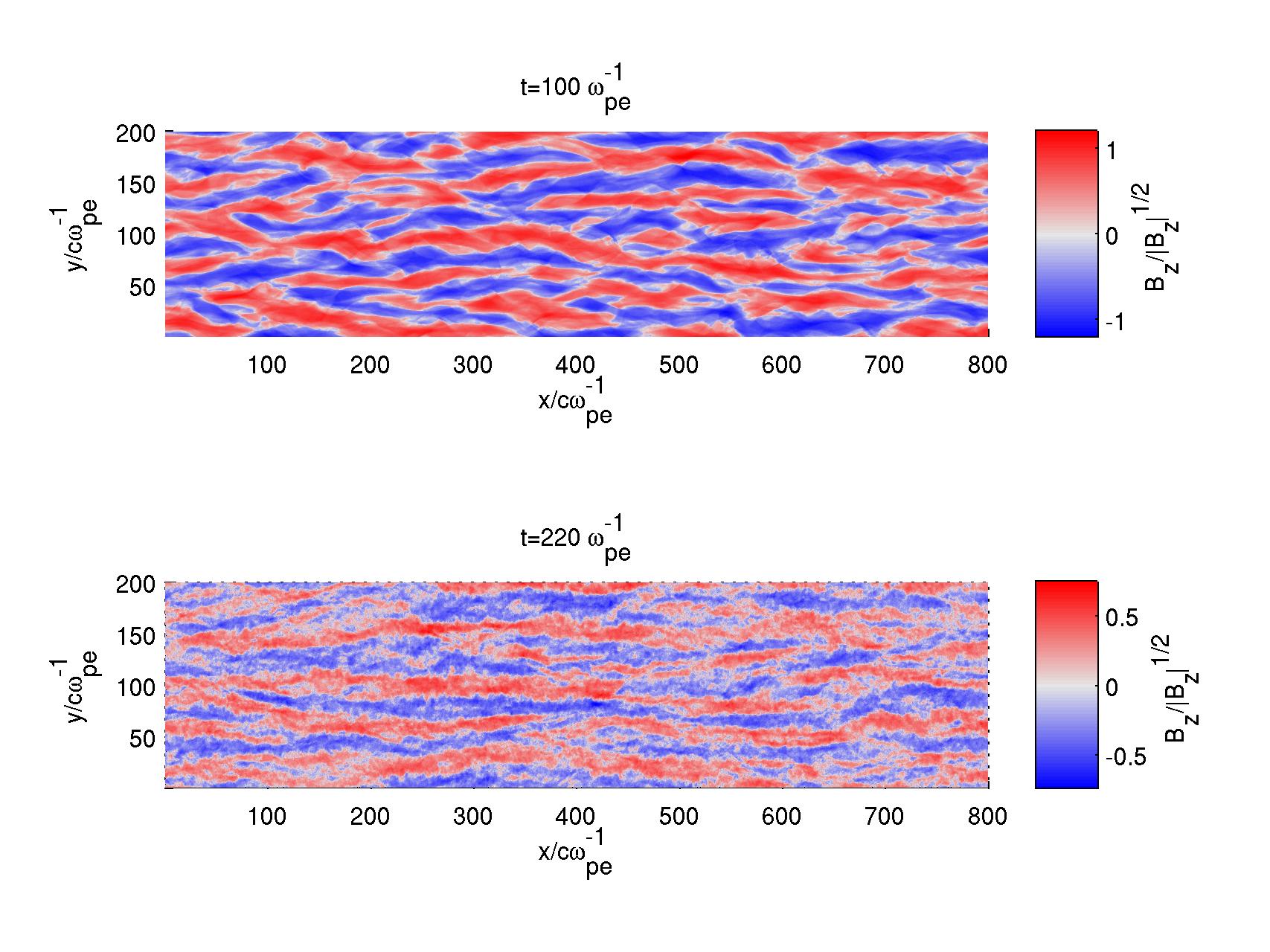}
    \caption{ The strength and sense of the magnetic field is shown in a small patch of the simulation box for two variant of simulations $M_{64}$ when transverse size if the filaments is about 30 $c/\omega_{pe}^{-1}$ . Top panel : only counter-streaming ions beams (immobile or infinite mass electrons are to be assumed to be sitting at the initial locations of ions in order to achieve charge neutrality). Bottom panel : simulation $M_{64}$ in which case electrons are also participating in the plasma dynamics. Time such that the transverse size of the filaments is about 40 $c/\omega_{pe}$. All modes grow faster in the absence of electrons which resonantly suppress the growth of waves.}
\label{fld_compare}
\end{figure}

The role of electrons in suppressing the oblique modes can clearly be demonstrated by simulating the Weibel instability with electrons of infinite mass such that electrons do not interact with the plasma waves.  In the absence of lighter resonating electrons all modes grow faster and the Weibel instability give rise to rather stronger magnetic field (figure \ref{fld_compare}). In case of infinitely massive electrons the current filaments which form due to the Weibel instability are rather short in length along the longitudinal direction and get quickly disoriented as compared to the simulation with electrons interacting with the waves.

The large fluctuation in the amplitudes of the waves (figure \ref{oblique}) is apparently due to interference of waves with same wave number but with different phases appearing in the plasma at different spatial locations which grow independently. The phase difference in the purely transverse Weibel modes at different locations can also be blamed for the wiggle in the filaments which appears at the very beginning of the simulation and can contribute to the scattering of the electrons. 

The power spectrum of the electromagnetic fields (Figure \ref{power_spectrum}) reveals a broadband nature of the instabilities in the counter-streaming plasma. We have identified the growth of transverse and oblique Weibel-like modes as well as parallel electrostatic modes in our simulation.  However, there can possibly be several other waves and instabilities in the counter-streaming plasma \citep{Bret2009}, which make the current filaments unstable, and might become more pronounced for plasma parameters which are rather more realistic for astrophysical contexts, i.e. for higher ion to electrons mass ratio or larger Lorentz factor of the streaming plasma than simulated here. In any case, it is evident that whichever wave can bend the current filaments of streaming ions can very efficiently scatter electrons out of the filaments since the electrons have much less inertial mass as compared with ions. The bending of the filaments, which we suggest is due to a broadband nature of the Weibel instability (i.e. growth of the waves with their wave vector at some angle with respect to the transverse direction), is a crucial requirement for the heating of the electrons as demonstrated by the one-dimensional simulation where poor scattering results in trapped electrons, which efficiently short out the inductive electric field created by streaming ions and do not get energized.

\subsection{One dimensional simulation: purely transverse Weibel mode}
In order to illustrate the role of oblique and electrostatic modes parallel to the streaming direction we simulated the counter-streaming plasma in one dimension such that the only modes transverse to the streaming direction, i.e. purely transverse Weibel models, are allowed to grow. The growth of oblique as well as longitudinal modes is suppressed by reducing the size of the box along the streaming direction (x-axis) to sub-skin depth level, hence reducing the simulation to effectively 1.5 dimension. That is to say, though electromagnetic field and particle's velocity are allowed to have all three components, they are allowed to vary along the transverse direction (y-axis) only. Here we discuss results from the simulation which has two dimensional box of size 0.3 $c/\omega_{pe} \times$ 1600 $c/\omega_{pe}$ and 64 particles per cell. All other parameters for this simulation are same as in the case of two dimensional simulation $M_{64}$.
\begin{figure}
    \centering
    \includegraphics[width=0.5\textwidth]{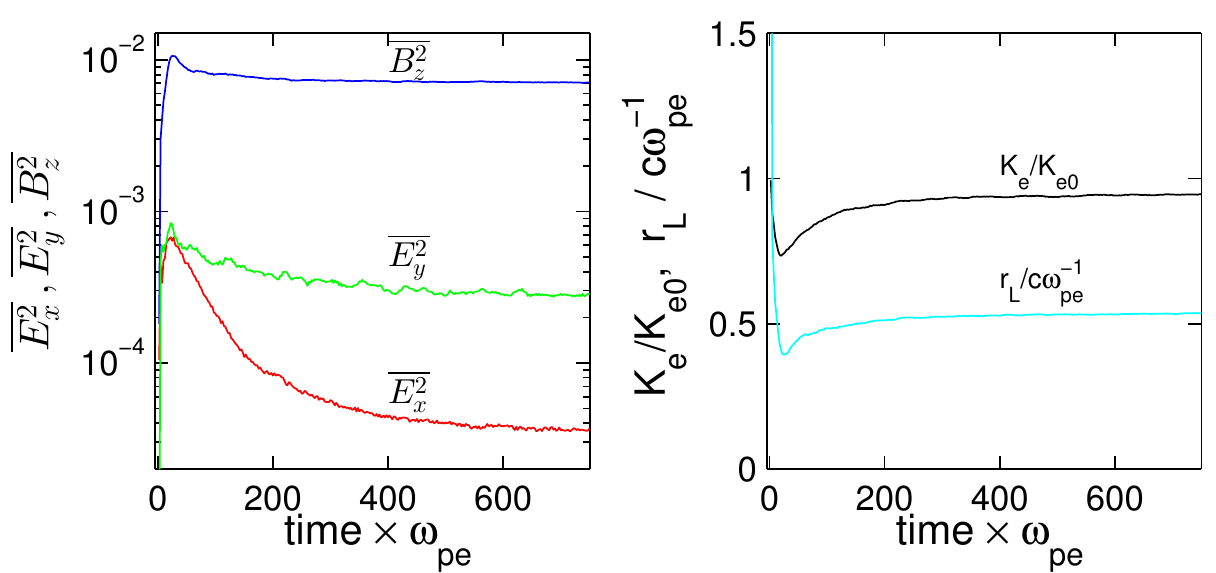}
    \caption{One dimensional simulation: A) Left panel: Temporal variation of the mean square (averaged over the physical domain of the simulation) of the electromagnetic field components $E_x$ (red), $E_y$ (green), and $B_z$ (blue) are normalized to the total initial kinetic energy density $n(m_i+m_e)(\gamma_0-1)c^2$. B) Right Panel : a) Black curve : Temporal variation of Kinetic energy of electron $KE_e$ (normalized to the initial kinetic energy of electrons $KE_{e0}=(\gamma_0-1)m_ec^2$),  b)  Cyan Curve: Ratio of instantaneous Larmor radius to the instantaneous skin depth of electrons as in the figure \ref{rlsd}. }
    \label{oned1}
\end{figure}

In the one dimensional simulation electrons first undergo the Weibel instability, which essentially is the Weibel instability of one species. The electron Weibel stage ends with the isotropisation of electrons.  As the ions undergo the Webiel instability size of ion current filaments and strength of the magnetic field grow. At the stage when transverse size of the current filaments becomes $\sim 2\pi c/\omega_{pe}$ electrons, which are magnetized, get confined in ion current filaments between two adjacent peaks of the magnetic field \citep{Achterberg2004, Yuri2006}. Electrons in the current filaments quiver in transverse direction and drift in the same sense as the ions. The countercurrent due to electrons efficiently arrest the growth of Weibel instability of the ions. Anisotropic ions continue to stream in the current filaments and the growth of magnetic field ceases (Figure \ref{oned1}). The lack of heating of electrons observed in the one dimensional case can be attributed to the lack of scattering of electrons out of the current filaments which is rather efficient in the two-dimensional case due to obliques waves growing along with the purely transverse mode.  

\section{Discussion and Conclusions}
We have simulated the development of the Weibel instability in relativistically counter-streaming homogenous ion-electron beams, using the kinematic PIC method. The physical domain of the simulation is taken to be sufficiently large, in both parallel and perpendicular to the plasma streaming direction, to ensure that the growth of Weibel unstable modes as well as the longitudinal electrostatic states are not frustrated due to finite size of the box. The homogeneous set-up simulated here ensures that any effect of large scale longitudinal  inhomogeneity which may be present in the case of shock is separated out. In the case of a shock transition there might be some additional heating due to cross-shock potential that may develop because of longitudinal separation of ions from electrons, since the lighter electrons can be relatively easily isotropised by the foreshock magnetic field, hence interrupting the electrons flow before the flow of ions \citep{Gedalin1993,Yuri2006b,Yuri2006}.  

Our findings from the plasma simulations concerning the heating of electrons can be summarized as follows. 

\begin{enumerate}
\item In the case of relativistic counter-streaming homogeneous plasma beams electrons are accelerated to the energy comparable to the energy of ions and most of the heating occurs during the formation and bending of the filaments and little after the filaments disrupt and disorient.
\item Comparison of the net work done by longitudinal and transverse electric field shows that the main force that takes energy aways from the ions and accelerates electrons is due to the longitudinal electric field. 
\item The work done by the transverse electric field is negligibly small as compared to the the longitudinal electric field and increasingly so for larger ions to electrons mass ratios (figure \ref{edotv}).

\item   Decomposition of the electric field into electrostatic and inductive component reveals that the longitudinal components is partially inductive, which is due to the growing current in the current filaments, and partially electrostatic, which is due to bending of the current filaments and growth of electrostatic waves along the longitudinal direction. The transverse component, on the other hand, is mainly electrostatic in nature and is due to the separation of charges. 

\item Calculation of rate of net work done on the electrons and ions by the electrostatic and inductive components of the longitudinal electric field shows that the work done by the both components are comparable, with that of the inductive component being slightly larger. It suggests that part of the acceleration of electrons is due to the fast moving chunks of current filaments (fermi-like heating, presumably second order). 

\item We showed that background electrons efficiently take energy away from the waves generated by the unstable counter-streaming ions and hence significantly alter the dynamics of the instabilities in counter-streaming plasma. It then becomes essential to take continuous energization of the electrons into account in order to calculate growth of various wave modes, even in the linear approximations. 

\item Fourier spectrum of the electromagnetic fields suggest that waves of same wavenumber but with different phases develop in the simulation and that there is significant phase mixing which can partially be responsible for the instability of filaments and scattering of electrons.

\end{enumerate}

   Current filamentation and significant electron heating is observed in three dimensional simulations of countersteaming plasma as well. We expect that the physical mechanisms in a more realistic three dimensional simulation should be the same as in the two dimensional case discussed here and can be verified in future simulations. 

We thank U. Keshet, Y. Lyubarsky, and A. Spitkovsky for helpful discussions. We thank A. Spitkovsky for a critical reading of the manuscript. We are grateful to U. Keshet for kindly providing us with computational resources for the simulations. RK and DE acknowledge support from the Israel-U.S. Binational Science Foundation and the Israeli Science Foundation.  
\bibliography{elc_heating}
\end{document}